\documentclass[aps,prl,twocolumn,superscriptaddress,a4paper]{revtex4}

\usepackage{graphicx}

\pdfoutput=1 %Needed for ArXiv submission

\usepackage{fancyhdr}
\usepackage[colorlinks=true,citecolor=blue,linkcolor=magenta]{hyperref}
\hypersetup{%
pdfauthor = {Patrick Maletinsky},
colorlinks = true, linkcolor = blue, urlcolor=blue, bookmarksnumbered =  true}

\begin{document}

%\preprint{}

\title{Coherent, mechanical control of a single electronic spin}


\author{S. Hong\footnote{These authors contributed equally to this work}}
\affiliation{School of Engineering and Applied Science, Harvard University, Cambridge, Massachusetts, 02138 USA}
\author{M.S. Grinolds\footnotemark[\value{footnote}]}
\affiliation{Department of Physics, Harvard University, Cambridge, Massachusetts 02138 USA}
%\email[]{grinolds@fas.harvard.edu}
\author{P. Maletinsky\footnotemark[\value{footnote}]}
\affiliation{Department of Physics, Harvard University, Cambridge, Massachusetts 02138 USA}
%\email[]{patrickm@physics.harvard.edu}
\author{R.L. Walsworth}
\affiliation{Harvard-Smithsonian Center for Astrophysics, Cambridge, Massachusetts 02138 USA}
\author{M.D. Lukin}
\affiliation{Department of Physics, Harvard University, Cambridge, Massachusetts 02138 USA}
\author{A. Yacoby}
\affiliation{Department of Physics, Harvard University, Cambridge, Massachusetts 02138 USA}
\email[]{yacoby@physics.harvard.edu}
%\thanks{}

\date{\today}

\begin{abstract}
The ability to control and manipulate spins via electrical\,\cite{Hanson2007,Nowack2007,Foletti2009}, magnetic\,\cite{Poole1983,Jelezko2004} and optical\cite{Press2008} means has generated numerous applications in metrology\cite{Chernobrod2005} and quantum information science\cite{Nielsen2000} in recent years. A promising alternative method for spin manipulation is the use of mechanical motion, where the oscillation of a mechanical resonator can be magnetically coupled to a spinÕs magnetic dipole, which could enable scalable quantum information architectures9 and sensitive nanoscale magnetometry\cite{Balasubramanian2008,Maze2008,Taylor2008}. To date, however, only population control of spins has been realized via classical motion of a mechanical resonator\cite{Rugar2004,Mamin2007,Wang2006}. Here, we demonstrate coherent mechanical control of an individual spin under ambient conditions using the driven motion of a mechanical resonator that is magnetically coupled to the electronic spin of a single nitrogen-vacancy (NV) color center in diamond. Coherent control of this hybrid mechanical/spin system is achieved by synchronizing pulsed spin-addressing protocols (involving optical and radiofrequency fields) to the motion of the driven oscillator, which allows coherent mechanical manipulation of both the population and phase of the spin via motion-induced Zeeman shifts of the NV spinÕs energy. We demonstrate applications of this coherent mechanical spin-control technique to sensitive nanoscale scanning magnetometry.
\end{abstract}

\maketitle

\begin{figure}
\includegraphics[scale=1]{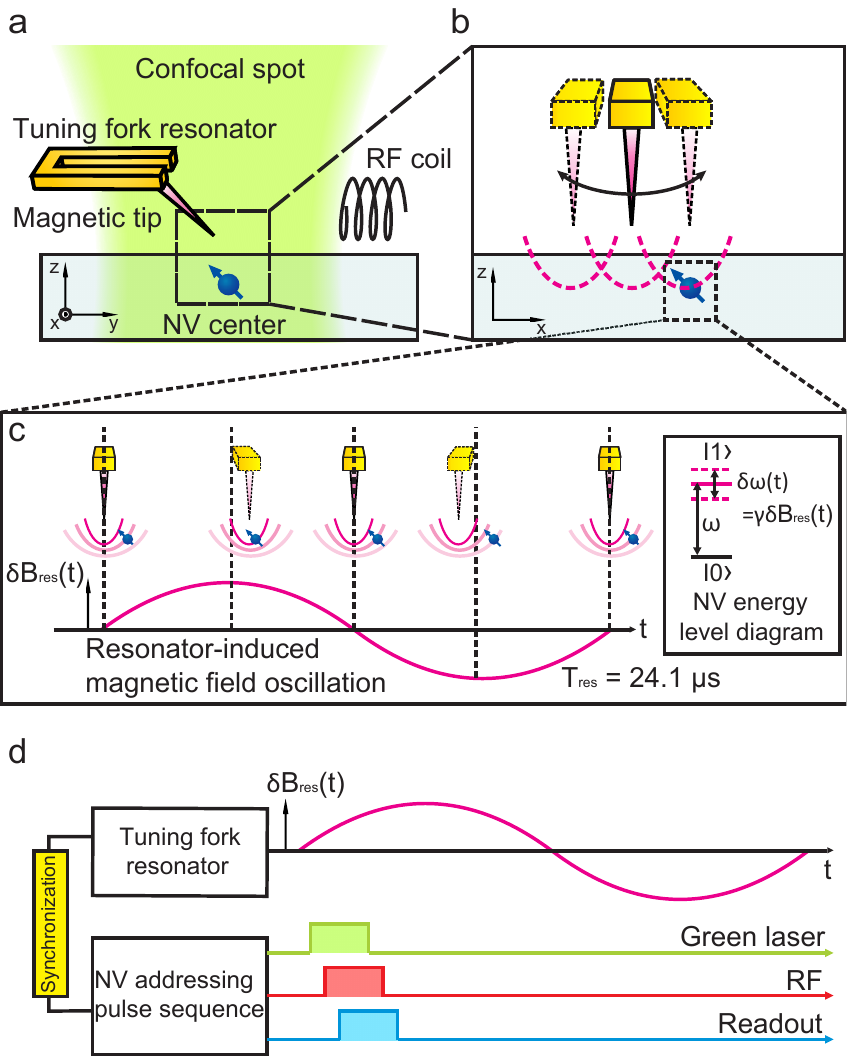}
\caption{\label{Fig1} Coherent dynamics of a mechanical resonator and an electron spin. (a) Schematic of an electron spin of an NV center next to a tuning fork resonator. A magnetic tip provides coupling between the tuning fork resonator and the NV spin. Readout and addressing of the NV spin is provided through a confocal microscope and an RF coil. The details of the experimental setup are described in Grinolds et al.\,\cite{Grinolds2011}. (b, c) Coupling of the motion of the resonator to the electronic spin of the NV center. The local magnetic field at the position of the NV spin changes as a function of the resonator's motion due to the magnetic field gradient of the tip. This resonator-induced magnetic field oscillation, $\delta B_{\rm res}(t)$, with the oscillation period, $T_{\rm res}=24.1~$µs, modulates spinÕs dipole transition energy, $\delta\omega(t)$, through a Zeeman shift, $\delta\omega(t)=\gamma\delta B_{\rm res}(t)$, where $\gamma$ is the gyromagnetic ratio of the spin (Inset). (d) Scheme for coherent coupling of the resonatorÕs motion to the NV spin. The resonatorÕs motion is synchronized to the NV addressing sequences. After this synchronization, the motion of the resonator influences the NV spin coherently with respect to the standard optical/RF pulse sequences which addresses the NV spin. }
\end{figure}

The magnetic coupling between spins and mechanical resonators has been recently investigated for imaging the locations of spins via magnetic resonance force microscopy\cite{Rugar2004,Mamin2007}. as well as for sensing nanomechanical resonator motion\cite{Arcizet2011}. A scrarcly explored resource in such coupled spin-resonator systems is the coherence of a driven resonatorÕs motion, and here, we demonstrate that this motion can be used to fully control the quantum state of an individual electron spin. We employ an electronic spin associated with an NV center in diamond as the target spin due to the efficient optical initialization and readout of the NV spin\,\cite{Gruber1997}, as well as its long coherence time\,\cite{Balasubramanian2009}. A single NV center is prepared near a bulk diamond surface (at a nominal depth of $10~$nm, see Methods), and its spin-state is read out optically through spin-dependent fluorescence\,\cite{Gruber1997}. The mechanical resonator is a quartz tuning fork, with a micro-fabricated quartz tip attached at the end of one of its prongs. The tuning fork operates in a transverse oscillation mode with a resonant frequency of $41.53~$kHz and oscillation amplitude that can be controllably varied up to $250~$nm (see Fig.\,\ref{Fig1} and Methods). Magnetic coupling between the NV spin and the tuning fork resonator is provided by a $25~$nm CoFe magnetic film evaporated onto the apex of the quartz tip (see Methods). Consequently, the transverse mechanical motion of the tip generates an oscillatory magnetic field at the spinÕs location, which modulates the Zeeman splitting between the NVÕs energy levels (Fig.\,\ref{Fig1}b, c). To achieve coherent control of the target NV spin using the resonatorÕs mechanical motion, we synchronize spin manipulation protocols to the driven oscillation of the mechanical resonator (Fig.\,\ref{Fig1}d). With this synchronization, the relative timing between applied radiofrequency (RF) pulses and the resonator motion is fixed, allowing for the resonator's motion to coherently and deterministically influence the NV spin state. 

\begin{figure}
\includegraphics[scale=1]{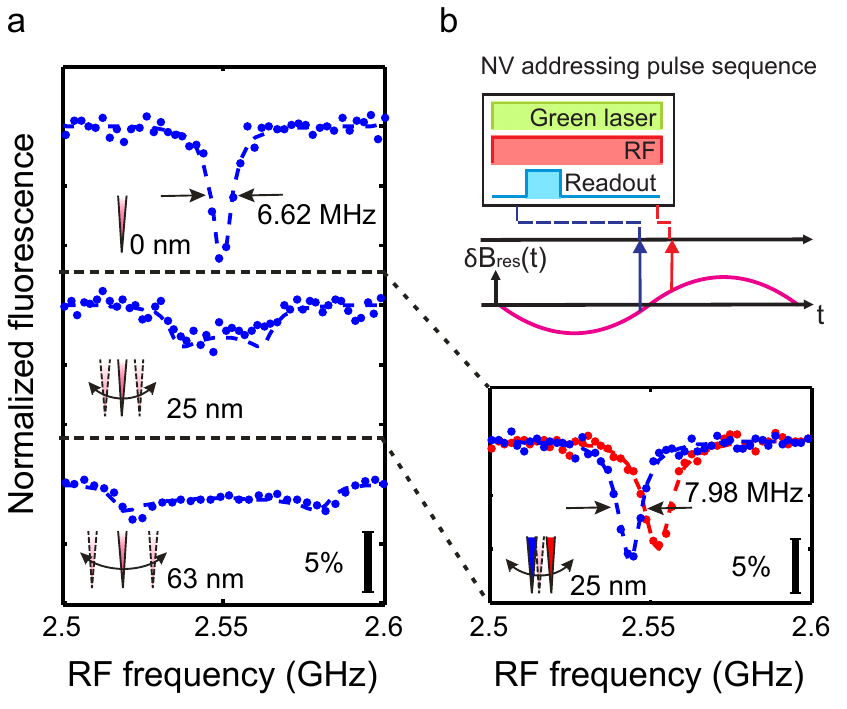}
\caption{\label{Fig2} Spin-resonator coupling as observed by electron spin resonance (ESR). (a) Motion-induced broadening of ESR. As the amplitude of the resonatorÕs oscillation is increased, the ESR linewidth broadens beyond its initial value ($6.62\pm0.49~$MHz, here power-broadened by the applied RF field). The shape of ESR evolves to a bimodal form, owing to the simple harmonic oscillation of the resonator. %(see Supplementary Information). 
(b) Stroboscopic ESR measurement. Measurements are taken synchronously to the motion of the resonator with $1~\mu$s acquisition time. Having this short acquisition time compared to the oscillation period of the resonator ($24.1~\mu$s) results in stroboscopic ESR snapshots, which capture the applied tip field - and the corresponding shift in ESR - as a function of the resonatorÕs position relative to the spin. The actual data (blue and red) are taken with the timings of $-0.8~\mu$s for the blue and $1.8~\mu$s for the red relative to the node of the resonatorÕs oscillation. The linewidth of the blue colored ESR is $7.98\pm0.43~$MHz, which is nearly recovered to the original linewidth of $6.62~$MHz.}
\end{figure}

The pronounced coupling between the resonatorÕs mechanical motion and the NV spin is measured with optically detected electron spin resonance (ESR). The ESR of the target NV center is acquired by sweeping the frequency of an applied RF field through the NV spinÕs magnetic dipole transition and collecting the resulting NVÕs spin-state-dependent fluorescence, with lower (higher) fluorescence when RF field is on (off) resonance with the NV spin transition\,\cite{Gruber1997}. As the oscillation amplitude of the tuning fork increases, we observe a broadening in the ESR spectrum (Fig.\,\ref{Fig2}a), resulting from the larger magnetic field modulation associated with the greater range of motion of the resonatorÕs magnetic tip. As the driving amplitude of the resonator is further increased, the ESR lineshape becomes bimodal, which reveals the distribution of dwell-times of the magnetic tip as a function of its position during the oscillatory motion, with the two peaks representing the turning points of the oscillation trajectory. %(see Supplementary Information). 
 To deconvolve this time-averaged spectral broadening in the coupled dynamics of the NV spin and resonator, we perform stroboscopic ESR measurements synchronized to the resonatorÕs motion (Fig.\,\ref{Fig2}b), which reveal the magnetic tipÕs position as a function of time via the tip-induced Zeeman shift of the NVÕs ESR resonance frequency. Figure\,\ref{Fig2}b shows two example stroboscopic ESR spectra with shifted resonance frequencies consistent with the magnetic tipÕs position in its oscillatory cycle. The observed remaining broadening of the ESR linewidth (full width at half maximum) is within $20\%$ of the value we find for an undriven magnetic tip ($6.62\pm0.49~$MHz). Such measurements are acquired with an acquisition time ($1~\mu$s) much shorter than the resonatorÕs oscillation period ($24.1~\mu$s), which enables stroboscopic readout of the NV spinÕs resonance frequency for a well defined position of the magnetic tip. The near-restoration of the zero-tip-motion ESR linewidth at arbitrary tip position confirms that the resonator motion is coherent with the spin addressing protocols.

\begin{figure}
\includegraphics[scale=1]{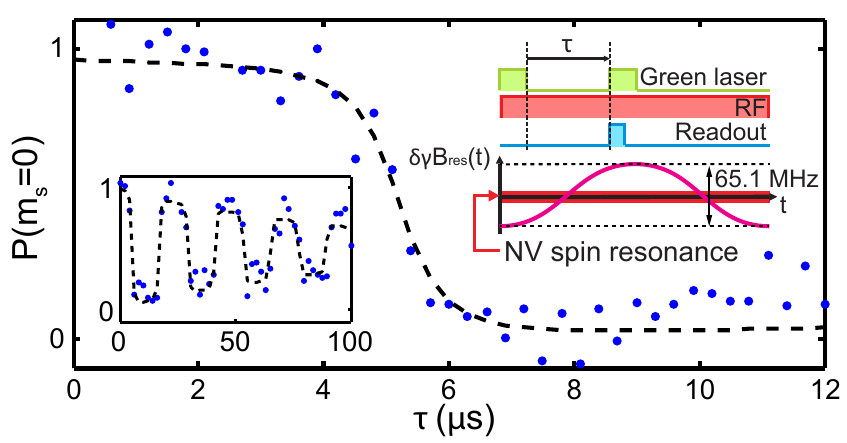}
\caption{\label{Fig3} Population control of a spin with the mechanical motion of a resonator. The motion of the resonator periodically modulates the local magnetic field at the position of the target spin, and therefore modulates its spin resonance. Adiabatic spin inversions occur when the resonator sweeps the spin resonance through the frequency of an applied RF field (inset, right). The corresponding range of frequency sweep is $65.1~$MHz, larger enough than RF Rabi frequency, $6.25~$MHz to achieve high fidelity of spin inversion. Plotted is the population of the NV spin in the $m_s=0$ state ($P(m_s=0)$) as a function of delay between initialization and readout ($\tau$). The left inset shows a periodic population modulation with multiple cycles and a period of $24.1~\mu$s in accordance with the measured resonator frequency. Fits to the adiabatic passages take into account a finite RF Rabi frequency of $6.25~$MHz. %(see Supplementary Information). 
The decay in the inversion amplitude is determined by the spin-lifetime, $T_1$.}
\end{figure}

This synchronization of the resonatorÕs motion with respect to the external optical/RF control can be used for coherent control of both the population and phase of the target spin states. Population control (Fig.\,\ref{Fig3}) is achieved via an adiabatic fast passage\,\cite{Slichter1990}. The oscillation of the magnetic field induced by the magnetic tip motion modulates the NV ESR frequency. By fixing the applied RF frequency ($\omega_{\rm RF}$) to the center of this frequency modulation range, the target spinÕs population is adiabatically inverted each time the resonator-induced magnetic field brings the NV ESR frequency onto resonance with $\omega_{\rm RF}$. This spin population inversion occurs twice for each period of the resonator oscillation (here $24.1~\mu$s), resulting in periodic population inversion (i.e. spin flips). Note that such adiabatic population inversion can have a higher fidelity than conventional RF $\pi$-pulses, since it is robust against inhomogeneous variation in ESR frequency. It is especially useful for flipping ensembles of spins, where differing local environments for individual spins can otherwise limit control fidelity.

\begin{figure}
\includegraphics[scale=1]{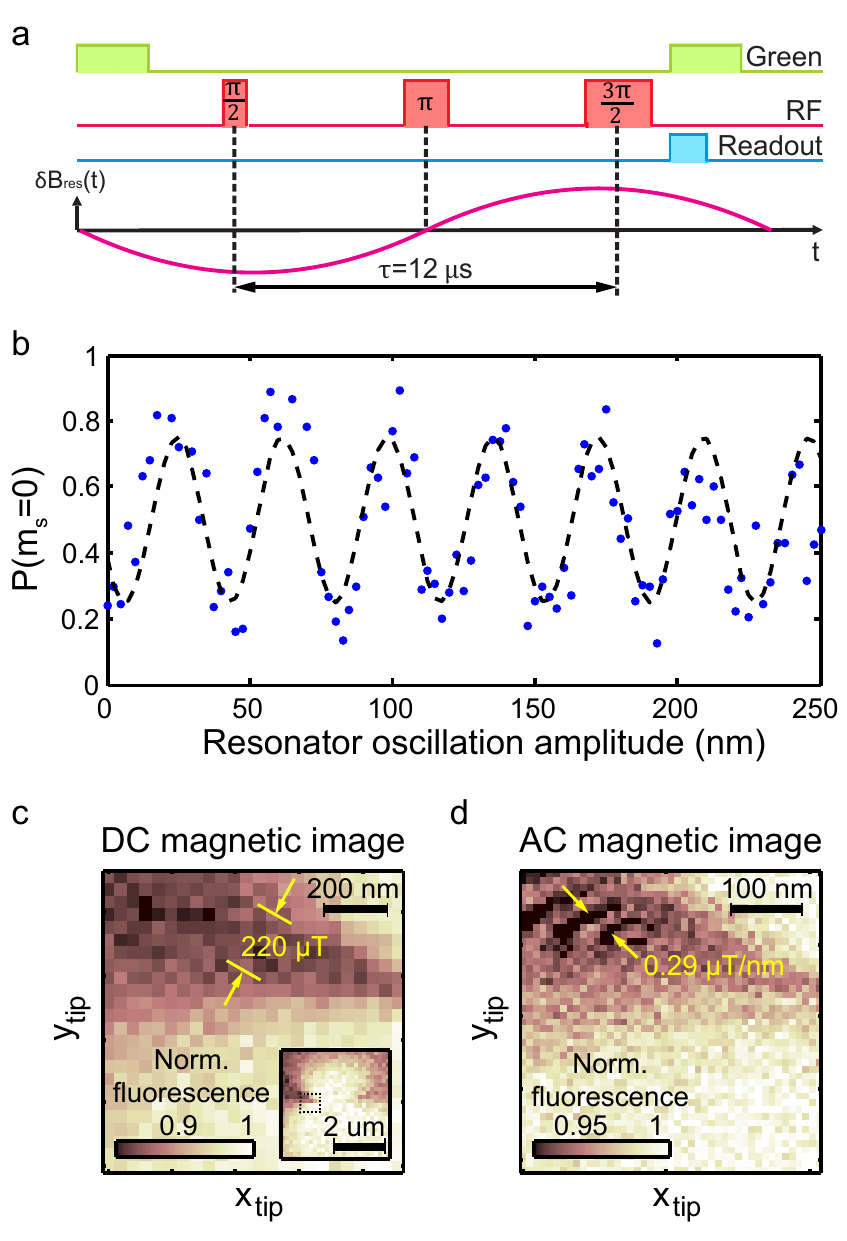}
\caption{\label{Fig4} Phase control of a spin with the mechanical motion of a resonator and applications to magnetic imaging. (a) Phase accumulation scheme based on a Hahn echo sequence. The evolution time, $\tau$, is fixed to $12~\mu$s in this experiment. The echo sequence is symmetrically placed at the node of resonatorÕs oscillatory motion, allowing for robust phase accumulation, which is only sensitive to the motion of the resonator, while canceling out slowly varying background magnetic fluctuations. (b) Phase accumulation of a spin as a function of the oscillation amplitude of the resonator.  The accumulated phase is proportional to the magnetic field modulation induced by the resonator, which scales linearly with the oscillation amplitude of the resonator. When the phase is converted to the population of the NV spin in the $m_s=0$ state ($P(m_s=0)$) by a RF $3\pi/2$-pulse, we observe a sinusoidal oscillation of the population. (c) Scanning DC magnetometry with the NV spin. The external RF frequency is set on resonance with ESR in the absence of the magnetic target. The change in local magnetic field at the position of NV results in ESR frequency shifts, which then changes the spin-state-dependend fluorescence. The larger scan image in the inset reveals a dipole-like pattern of the tip-induced magnetic field. The scan is zoomed to a region depicted by the dotted square box. The darker region is the region where the spin is near resonance with the external RF frequency. (d) Scanning AC magnetometry with the NV spin. The magnetic tip is scanned laterally with $20~$nm oscillation amplitude. The modulation of the NV position relative to the magnetic target (here, the magnetic tip) produces a modulation of the magnetic signal from the source, providing a general method for adapting optimally sensitive AC magnetic imaging scheme. In comparison to DC magnetic image, we observe additional features, multiple fringes, which correspond to contours of constant magnetic field-gradient along the direction of the resonatorÕs oscillation.}
\end{figure}

Furthermore, synchronizing the resonator motion with an appropriate RF pulse sequence allows for high fidelity control of the target NV spinÕs phase. As a demonstration of such phase control, the NV spin is first prepared in a superposition state, using optical pumping followed by an RF $\pi/2$-pulse. Next, the time-varying magnetic field induced by the resonatorÕs motion results in a detuning of the NV ESR frequency from $\omega_{\rm RF}$, causing the spin to acquire a differential phase relative to its precession at $\omega_{\rm RF}$. In order to accumulate the phase with high fidelity, and to minimize the effect of other, incoherent magnetic field fluctuations, a Hahn echo sequence is employed and synchronized to the driven motion of the resonator (Fig.\,\ref{Fig4}a). By placing the pulse-sequence symmetrically with respect to the node of the resonator oscillation, the acquired phase is maximized for a given tip amplitude. For a fixed duration of the echo sequence (here, $\tau=12~\mu$s), the amount of accumulated phase is proportional to the amplitude of the magnetic field oscillation, which is given by the oscillation amplitude of the magnetic tip displacement multiplied by the tip-induced magnetic field gradient along the direction of the tip motion. To read out the accumulated phase, a final RF pulse (here, $3\pi/2$) is applied to project the accumulated phase of the NVÕs spin onto a distribution of the state populations, which is then measured via spin-state-dependent fluorescence. Figure\,\ref{Fig4}b shows an example of such resonator-induced spin phase accumulations, controlled by the amplitude of the resonatorÕs oscillation.
In these demonstrations of coherent, mechanical control of a single NV spin, the rate of spin manipulation is set by the $41.53~$kHz resonance frequency of the quartz tuning fork. However, nano-mechanical resonators can have resonance frequencies ranging up to nearly $1~$GHz\,\cite{Li2007,Liu2008}, which would allow rapid control of target spins. For example, a resonator with $1~$MHz frequency could perform more than $1000$ coherent spin manipulations within the demonstrated NV spin coherence time of a few milliseconds\,\cite{Balasubramanian2009}.

With the demonstrated population and phase control, we have achieved coherent, mechanical control over the state of a single electron spin. This coherent mechanical spin-control technique enables new applications in sensitive, nanoscale metrology. In particular, electronic spins in NV centers have been recently identified as excellent magnetic field sensors because of their long coherence time and efficient optical readout\,\cite{Balasubramanian2008,Maze2008,Taylor2008}, even under ambient conditions. Previous work\,\cite{Maze2008,Taylor2008} showed that optimal magnetic field sensitivity is achieved when the target field is modulated with a period comparable to the NVÕs spin coherence time (ÒAC magnetometryÓ). However, such AC magnetometry cannot be applied a priori to targets with static magnetizations. Our coherent mechanical spin-control scheme provides a solution as the motion of the tip transforms a spatially varying, static magnetic field of a magnetic sample into a time-varying magnetic field at the position of the NV center. We demonstrate such mechanical-resonator-enabled AC magnetometry by performing scanning, nanoscale magnetic field imaging of our magnetized tip (Fig.\,\ref{Fig4}c,d). At first, we acquire a ÔDC magnetic imageÕ\,\cite{Balasubramanian2008,Grinolds2011} by scanning the magnetic tip laterally near a single NV center, where the external RF frequency is fixed on resonance with NV ESR frequency in the absence of the tip, and the fluorescence change due to Zeeman shifts is monitored (Fig.\,\ref{Fig4}c). In this mode, the change in signal directly reflects the change in the local magnetic field at the position of NV. Consequently, a resonance band with decreased fluorescence is formed where the magnetic field associated with the tip brings NV ESR frequency on resonance with the external RF frequency. From this measurement, we extract a DC magnetic field sensitivity of $45~\mu$T$/\sqrt{\rm Hz}$. Next, we perform resonator-motion-enabled AC magnetometry (Fig.\,\ref{Fig4}d), where the same optical/RF-pulse sequence is used as for the phase-control experiment described above, while the driven motion of the magnetic tip is synchronized to the NV addressing protocols with fixed tip oscillation amplitude of $20~$nm. We calculate our NV spinÕs AC magnetic field sensitivity to be $0.92~\mu$T$/\sqrt{\rm Hz}$,% (see Supplementary Information), 
which constitutes a factor of $50$ improvement over our DC experiment. In contrast to the ÔDC magnetic imageÕ (Fig.\,\ref{Fig4}c), additional structures in the form of multiple interference fringes in the resonance region are revealed. In this configuration, the NV center senses magnetic field variations along the direction of the tip oscillation, and the observed fringes (Fig.\,\ref{Fig4}d) correspond to contours of constant magnetic field gradients, with neighboring fringes differing by a gradient of $0.29~\mu$T/nm.% (see Supplementary Information). 

A particularly appealing application of our motion-enabled AC magnetometry could be sensitive imaging of rapidly varying, but weak magnetic features, such as antiferromagnetically ordered systems. Our magnetic imaging technique optimizes magnetic field sensitivity through AC Magnetometry and should thereby be capable of detecting the magnetic moment of a single Bohr magneton within few seconds of data acquisition %(see Supplementary Information) 
Ð it therefore provides the sensitivity required to detect antiferromagnetic order with close to lattice-site resolution. While in our demonstration of motion-enabled AC magnetometry, the NV spin-sensor is located in a fixed bulk diamond sample, our scheme can be readily applied to studying arbitrary samples in a scanning geometry, if the NV spin-sensor is located on the tip of a scannable diamond structure such as a diamond nanopillar\,\cite{Maletinsky2012}.  
An additional application of our coherent mechanical spin-control technique is motion sensing for nanoscale mechanical resonators. Such detection of motion, while routinely performed for microscale mechanical resonators using optical interferometry, remains challenging for nanoscale objects. Our mechanical spin-control scheme employs a single, atomically localized NV spin, thereby allowing nanoscale displacement and motion sensing. In principle, our demonstrated phase control scheme can be used to measure the amplitude and the phase of the motion of a resonator of interest. Using the measured magnetic tip field gradient of $18.4~\mu$T/nm combined with the NVÕs AC field sensitivity, the oscillator amplitude sensitivity in our setup is estimated to be $49.8~$pm$/\sqrt{\rm Hz}$, which is already comparable to the sensitivity ($\approx10~$pm$/\sqrt{\rm Hz}$) achieved by optical inteferometry of sub-micron sized resonators\,\cite{Kouh2005}. Similarly, the same scheme can be applied to measure the phase of the resonatorÕs oscillatory motion, with an estimated sensitivity of $5.1~$mrad$/\sqrt{\rm Hz}$ at an oscillation amplitude of $10~$nm.

The demonstrated sensitivity for metrology applications can be further optimized through a variety of experimentally demonstrated improvements. Such schemes of improvements include extending NVÕs spin coherence time\,\cite{Balasubramanian2008,deLange2010}and enhancing photon collection efficiency\,\cite{Maletinsky2012,Babinec2010,Hadden2010}. Our displacement sensing scheme can also be improved by engineering higher magnetic field gradients. For example, by using state-of-the-art magnetic tips with a field gradient of $1~$mT/nm\,\cite{Mamin2007}, an NV spin with $T_2$ of few ms, and enhanced photon collection efficiencies, the displacement amplitude sensitivity of $22.8~$fm$/\sqrt{\rm Hz}$ could potentially be achieved. Finally, we note that the techniques demonstrated here could be integrated with related methods for scanning-field-gradient spin MRI\,\cite{Grinolds2011} and super-resolution optical imaging and coherent manipulation of proximal spins\,\cite{Maurer2010}, with an ultimate goal of combined magnetic field sensitivity and spatial resolution to achieve real-time atomic-scale mapping of individual electron and nuclear spins in physical and biological systems of interest.

%\section*{Acknowledgments}
We gratefully acknowledge G. Balasubramanian and P. R. Hemmer for fruitful technical discussions, as well as B. Hausmann and M. Loncar for instruction in the fabrication of NV center containing nanostructures. We acknowledge financial support from NIST and DARPA. S. H. acknowledges support from the Kwanjeong Scholarship Foundation for fellowship funding. M. S. G. is supported through fellowships from the Department of Defense (NDSEG program) and the NSF. P. M.  acknowledges support from the Swiss National Science Foundation. This work was carried out in part at the Center for Nanoscale Systems (CNS), a member of the National Nanotechnology Infrastructure Network (NNIN), which is supported by the National Science Foundation under NSF award no. ECS-0336765. CNS is part of Harvard University.

%\section*{Competing Interests}
%The authors declare that they have no competing financial interests.

%\section*{Correspondence}
Correspondence and requests for materials should be addressed to A.Y. (\href{mailto:yacoby@physics.harvard.edu}{yacoby@physics.harvard.edu})
\\
\\
\\

% Create the reference section using BibTeX: Bibliography appended in the end
\bibliographystyle{apsrev4-1}
\bibliography{Coherent_mechanical_control_of_a_single_electronic_spin}

%\newpage
%
\section*{NV center samples:}
Individual NV centers are fabricated by first implanting $^{15}$N ions\,\cite{Rabeau2006} ($6~$keV) into ultrapure diamond (Element Six, electronic grade, $<5~$ppb nitrogen), resulting in a layer of implanted nitrogen atoms nominally $10~$nm below diamondÕs surface\,\cite{Ziegler2010}. The nitrogen atoms pair with nearby vacancies to form NV centers during an annealing process performed in vacuum at $750~^\circ$C. Our implantation and annealing parameters yield an NV density corresponding to one center every 50-100 nanometers. To isolate single NV centers, we use electron beam lithography to define an etch mask (Dow Corning, XR-1541) consisting of an array of spots whose sizes are matched to contain, on average, one NV center pet spot. A reactive-ion etch\,\cite{Lee2008} is then performed to remove exposed diamond surfaces, resulting in an array of diamond cylinders, each containing roughly one NV. 

\subsection*{Quartz tuning fork:}
In our experiment, a commercially available quartz tuning fork resonator (DIGI-KEY, X801-ND) is used. Its resonance frequency and quality factor are $41.53~$kHz, and $1400$, respectively. The excitation of the tuning fork is performed by mechanically dithering the tuning fork using a piezoelectric actuator. Additionally, the response of the tuning fork is read out electrically through piezoelectric detection. 

\subsection*{Magnetic tips:}
Magnetic tips are created by evaporating a magnetic layer onto quartz tips of roughly $80~$nm in diameter, which are fabricated using a commercial laser-pulling system (Sutter Instrument Co., P-2000). We use a thermal evaporator to deposit a $25~$nm layer of cobalt-iron on the side of the pulled quartz tip, followed by additional deposition of a $5~$nm chrome layer, which serves as a protective capping layer.

\end{document}